\documentclass[aps,prl,superscriptaddress,twocolumn]{revtex4-2}
\usepackage{comment}
\usepackage[utf8]{inputenc}
\usepackage{amssymb}
\usepackage{amsmath}
\usepackage{amsfonts}
\usepackage{bm}
\usepackage{mathtools}
\usepackage{bbold}
\usepackage{upgreek}
\usepackage{xfrac}
\usepackage{soul}
\usepackage{bm}
\usepackage{xcolor}
\usepackage{url}
\usepackage{txfonts}
\usepackage{bbm}
\usepackage{dsfont}
\usepackage{physics}
\usepackage{svg}
\usepackage[nolist]{acronym}

\newcommand\eqbydef{\stackrel{\mathclap{\mbox{\footnotesize{def}}}}{=}}

\DeclareSymbolFont{matha}{OML}{txmi}{m}{it}
\DeclareMathSymbol{\varv}{\mathord}{matha}{118}

\begin{document}

\title{Quantum Iterative Methods for Solving Differential Equations \\ with Application to Computational Fluid Dynamics}

\author{Chelsea A. Williams}
\affiliation{Department of Physics and Astronomy, University of Exeter, Stocker Road, Exeter EX4 4QL, United Kingdom}
\affiliation{PASQAL, 7 Rue Léonard de Vinci, 91300 Massy, France}
\author{Antonio A. Gentile}
\affiliation{PASQAL, 7 Rue Léonard de Vinci, 91300 Massy, France}
\author{Vincent E. Elfving}
\affiliation{PASQAL, 7 Rue Léonard de Vinci, 91300 Massy, France}
\author{Daniel Berger}
\affiliation{Siemens AG, Gleiwitzer Str. 555, 90475 N\"urnberg, Germany}
\author{Oleksandr Kyriienko}
\affiliation{Department of Physics and Astronomy, University of Exeter, Stocker Road, Exeter EX4 4QL, United Kingdom}
\affiliation{PASQAL, 7 Rue Léonard de Vinci, 91300 Massy, France}
\date{\today}
\begin{abstract}
We propose quantum methods for solving differential equations that are based on a gradual improvement of the solution via an iterative process, and are targeted at applications in fluid dynamics. First, we implement the Jacobi iteration on a quantum register that utilizes a linear combination of unitaries (LCU) approach to store the trajectory information. Second, we extend quantum methods to Gauss-Seidel iterative methods. Additionally, we propose a quantum-suitable resolvent decomposition based on the Woodbury identity. From a technical perspective, we develop and utilize tools for the block encoding of specific matrices as well as their multiplication. We benchmark the approach on paradigmatic fluid dynamics problems. Our results stress that instead of inverting large matrices, one can program quantum computers to perform multigrid-type computations and leverage corresponding advances in scientific computing. 
\end{abstract}
\maketitle
\begin{acronym}
  \acro{CFD}{computational fluid dynamics}
  \acro{LCU}{linear combination of unitaries}
  \acro{PDE}{partial differential equation}
  \acro{QLSA}{quantum linear systems algorithm}
  \acro{QSP}{quantum signal processing}
  \acro{VQA}{variational quantum algorithm}
\end{acronym}

\section{I. Introduction}

Differential equations (DEs) represent a fundamental tool for describing many natural and engineered phenomena. These range from the growth of populations in biology to heat transfer in thermal engineering and failure diagnosis in fracture mechanics \cite{braun1983differential}. A field that arguably needs the most cutting-edge DE solvers and high-performance compute is \ac{CFD} \cite{cfd_textbook}. This is since it requires integrating Navier-Stokes equations, which are highly nonlinear DEs that govern dynamics of fluid flow \cite{anderson1995computational}. Here particular challenges include turbulence modeling \cite{wilcox_book,Duraisamy_annurev}, predicting aerodynamics of aircrafts \cite{Strawn2002}, and simulating plasma physics for fusion (TOKAMAKs) \cite{Meyer2017} and astrophysics \cite{Browning2008}. The major hurdle for high-performing CFD simulations is the need to capture the multi-scale nature of phenomena \cite{Orszag1972,Hussain1983,Volker2005,Koumoutsakos_annurev} (boundary layer interactions, vortices etc.), requiring accurate fine resolution calculations with a cost that increases exponentially in dimensionality. Advancing algorithms for solving DEs that are inherently scalable and parallelizable is essential to tackle these issues effectively.

The algorithms for solving differential equations can be tentatively put in several families \cite{Gear1981}, each contributing special capabilities when solving CFD problems. A major family of DE solvers corresponds to grid-based methods such as finite differencing and finite element methods \cite{anderson1995computational,Liu2022rev}, where physical space discretization leads to efficient derivative approximations \cite{BERGER1984484}. State-of-the-art techniques include high-order finite volume schemes that are useful for capturing discontinuities in compressible flows \cite{JIANG1996202}. Grid-based methods often require solving large systems of linear equations that require the inversion of large matrices \cite{sparse_matrix_collection}. To operate at scale, they need to use iterative solving techniques \cite{Saad_Yousef_2003} where inversion is approached via successive approximation steps (e.g. using conjugate gradient or generalized minimal residual methods \cite{Youcef1986}), and advanced multigrid methods can be used to accelerate convergence \cite{Wesseling_book,MCBRYAN1991,Yserentant_1993}. A second family of DE algorithms corresponds to spectral methods \cite{boyd2013chebyshev,hussaini1983spectral}, where sets of basis functions (Fourier series, Chebyshev polynomials etc.) and fast transforms are used to compute derivatives and represent DE solutions \cite{trefethen2019approximation}. When combined with reduced-order modeling \cite{Touze2021,VIZZACCARO2021} and dynamic mode decomposition \cite{SCHMID_2010,Schmid_annurev}, they can capture relevant features and transient behavior of complex flows. Third, a family of lattice Boltzmann methods \cite{SUCCI_rev,Chen_annurev} based on modeling particle collisions and streaming dynamics is used for simulating complex multiphase flows (for instance in combustion applications \cite{DiRienzo_2012}). Finally, recent developments in CFD solvers come from the intersection between the fields of machine learning \cite{HodgesCFDML} and quantum information \cite{Balducci2022rev}. For the former, physics-informed neural networks combine a representative power of deep neural networks and automatic differentiation to solve data-driven CFD problems \cite{Raissi2020Science,Brunton2020,Markidis2021pinns,Vinuesa2022}, with examples in flow prediction in complex geometries \cite{PhysRevFluids.2.034603}, turbulence modeling \cite{Srinivasan2019,Eivazi2022}, and improvement of aerodynamic designs \cite{fluids7020062}. For the latter, the quantum-inspired approaches based on tensor networks (e.g. tensor trains) \cite{Oseledets2011,GarciaRipoll2021quantuminspired,ErikaNuno2022} utilize hierarchical structures and compressed representation of physical states to excel in representing turbulent flows \cite{Gourianov2022,Kiffner2023}.
\begin{figure*}[t]
    \centering
    \includegraphics[width=1.0\linewidth]{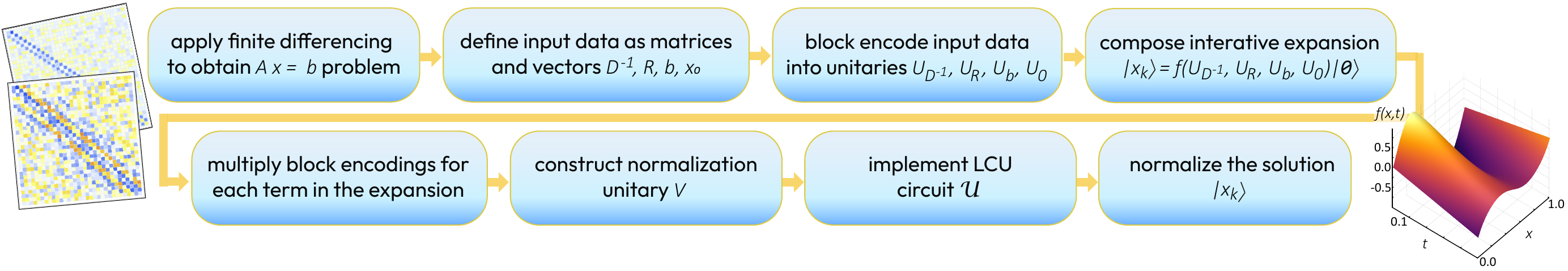}
    \caption{The algorithmic pipeline used to apply the quantum iterative methods via a state preparation protocol that is based on block encodings. The input is the differential equation and the output is the $k^\text{th}$ iterate solution. The steps are summarized in the boxes above, and details of each step are described in the text.}
    \label{fig:pipeline}
\end{figure*}


Quantum computing has emerged as a distinct approach to solve computational problems beyond the standard von Neumann architecture \cite{nielsen2010quantum}. Utilizing a state space that grows exponentially with the number of qubits and coherent evolution of entangled configurations, quantum algorithms offer promise in advancing DE algorithms and CFD solvers \cite{Succi_2023,Succi2024}. To date, many quantum DE algorithms represent analogs of grid-based protocols, where underlying advantage comes from solving linear systems of equations with improved scaling on the grid size \cite{hhl}. Here a \ac{QLSA} is used to perform the matrix inversion either via phase estimation \cite{hhl,Cao_2013,Berry_2014,Montanaro2016,ingelmann2023quantum,MartinSanz2023}, \ac{LCU} \cite{hamiltonian_simulation_lcu,childs2017quantum}, adiabatic protocols \cite{subacsi2019quantum}, or \ac{QSP} \cite{Lin2020optimalpolynomial,Linden2022quantumvsclassical,Krovi2023improvedquantum}. The optimal performance with condition number has been attained \cite{Costa2022}, and gate count estimates for large-scale modeling \cite{jennings2023efficient,costa2023improving,jennings2024cost} has set the goal for developing fault tolerant quantum computers. Another distinct set of DE algorithms corresponds to Schr\"{o}dingerization \cite{jin2022quantum,jin2022quantumEXT,jin2023analog,jin2023quantum,hu2024quantum}, where native evolution of quantum states is employed for efficient solution of time dynamical problems, and quantum reservoir computing \cite{Pfeffer2022,Pfeffer2023}. Also, a range of quantum lattice Boltzmann algorithms was proposed \cite{Mezzacapo2015,Succi2015,TODOROVA2020,Steijl2023,schalkers2022efficient,schalkers2023importance,Budinski2022,chrit2023fully,zamora2024efficient}, with potential advantages for high Reynolds number flows \cite{li2024potential}. Finally, quantum physics-informed DE solvers have been developed in the area of quantum scientific machine learning \cite{lubasch2020variational,kyriienko2021solving,Knudsen2020,Markidis2022qpinns,jaksch2022variational,mouton2023deep,paine2023physicsinformed}, targeting data-driven tasks \cite{heim2021quantum,Varsamopoulos2022,jaderberg2023let} and probabilistic modeling \cite{Paine2021,kyriienko2022protocols,kasture2022protocols}.

Comparing typical workflows used in \ac{CFD} and quantum DE solvers, we stress one conceptual difference. While many current quantum algorithms rely on the full problem matrix inversion (directly or effectively), this is atypical to CFD solvers. These solvers avoid inverting large matrices, due to memory and computational constraints \cite{sparse_matrix_collection}, and instead utilize iterative methods that are based on relaxation to obtain the approximate solution. Thus, we note the need for extending the quantum algorithmic toolbox such that iterative and multigrid methods can be accommodated \cite{Saad_Yousef_2003,Wesseling_book}. Recent research in this area includes proposals for the quantum Navier-Stokes equation solver with Kacewicz method and quantum amplitude amplification \cite{Gaitan2020,Oz2021,KACEWICZ2006676}, performing a gradient descent \cite{quantum_gradient_descent} (targeting applications in machine learning), and an analog to the Kaczmarz method based on qRAM \cite{quantum_kaczmarz_method}. Another approach considers generalized iteration schemes being embedded within a larger block linear system, to be solved with a \ac{QLSA} \cite{quantum_iterative_solver}. The iterative approaches were considered from the Schr\"{o}dingerization perspective and analog-type simulation in the continuous-time limit \cite{continuous_quantum_jacobi}. Additionally, a multigrid approach has been outlined in \cite{quantum_multigrid}, while not detailing the implementation.

In this paper we present a quantum algorithmic implementation of iterative differential equation solvers, aiming to build an end-to-end gate-based circuit construction. Specifically, we concentrate on the Jacobi method as an instructive example towards showcasing the core algorithmic elements for developing quantum-based iterative solvers. Different from prior art, we utilize a \ac{LCU} approach to generate an iterated state for the Jacobi method (Sec. II), and develop block encoding protocols specific to the method (also describing generalizations to other iterative schemes). This offers high-performance and flexibility in terms of allocated resources. We discuss the full details of the algorithmic implementation in Sec. III. Examples of applying the quantum Jacobi method to solve the Burgers equation and Euler's equations are presented in Sec. IV. The discussion in Sec. V highlights the resource requirements for the quantum method, identifying areas for further improvements, and considering the performance of the method in the context of \ac{CFD} applications. Conclusions are presented in Sec. VI.

\section{II. Algorithm}

\par The steps involved in constructing the algorithm for the quantum circuit implementation of quantum iterative methods are outlined in Fig.~\ref{fig:pipeline}. We start with the Jacobi method as an instructive example. Other methods follow a similar workflow, with a difference in steps corresponding to splitting matrix $A$ and designing block encodings for its parts. The ultimate aim of the proposed method is to iteratively solve a differential equation by off-loading the computationally intensive operations to a quantum processor. In this and the next section we detail the intuition and necessary tools needed for composing quantum iterative solvers.

\par The algorithm for iterative solving begins by applying finite differencing to the differential equation \cite{linear_systems_textbook}. This can also include a linearization step for nonlinear systems. The problem is then reduced to solving a system of $N$ equations specified by $A\vb*{x}=\vb*{b}$ with unknown solution $\vb*{x}$ and known right-hand vector $\vb*{b}$. For the Jacobi method the matrix $A=D+R$ is divided into a diagonal matrix $D$ and an off-diagonal matrix $R$. It is assumed throughout that $A$ is diagonally dominant to guarantee convergence. Instead of inverting $A$ directly, the Jacobi solver updates the solution incrementally to converge towards the final numerical approximation using the iterative scheme 
\begin{equation} \label{eq:classical_jacobi}
    \vb*{x}_k = D^{-1}(\vb*{b} - R\vb*{x}_{k-1}) \ \forall \ k\in[1,K].
\end{equation}
Note that the diagonal matrix $D$ is readily invertible for nonzero diagonal elements, greatly simplifying the process of obtaining the solution $\vb*{x}_k$. As the solver progresses from iteration $k-1$ to iteration $k$, the solver converges closer to the true value of $\vb*{x}$, starting from an initial guess $\vb*{x}_0$. After $K$ iteration steps, the resulting solution $\vb*{x}_{K}$ is obtained as a numerical approximation to the true solution.

\par The Jacobi iterative scheme in Eq.~\eqref{eq:classical_jacobi} can be recast as a quantum solver by mapping classical data (matrices) into unitary quantum circuits and utilizing post-selection. This involves identifying a protocol capable of loading the data. We achieve this by employing a block encoding technique, where any operator $O$ can be implemented as a projection of some unitary operator $U_O$ onto a tailored state $|G\rangle$, such that $O = (\langle G| \otimes \mathds{1}) U_O (|G\rangle \otimes \mathds{1})$ \cite{Low2019hamiltonian}. Here $\mathds{1}$ is the identity matrix acting on a part of the system. Specifically, we develop block encodings to embed $R$ into the unitary $U_R$, and $D^{-1}$ into the unitary $U_{D^{-1}}$ (discussed in details in the next section). In addition, the preparation of vectors $\vb*{x}_0$ and $\vb*{b}$ is achieved by constructing respective unitaries $U_0$ and $U_b$. The corresponding quantum states are then prepared as $\ket{x_0} = U_0\ket{\emptyset}$ and $\ket{b} = U_b \ket{\emptyset}$, starting from a fiducial quantum state $\ket{\emptyset}$ taken as a computational zero state.
%
%
The quantum Jacobi iterative algorithm can then be written as
\begin{equation} \label{eq:quantum_jacobi}
    \ket{x_k} = U_{D^{-1}} \bigl[U_b\ket{\emptyset} - U_R\ket{x_{k-1}}  \bigr],
\end{equation}
where the state $\ket{x_k}$ stores the iterate solutions. We detail the construction and implementation of the circuits used in Eq.~\eqref{eq:quantum_jacobi} in Sec. III. Later in Sec. IV, we apply the method to solve differential equations appearing in the field of fluid dynamics. We observe that the iterative approach is able to recover high-quality solutions in very few iterations, even for problems where the solution is discontinuous. This is particularly prominent in \ac{CFD} simulations where shock waves appear. 

\par The protocols and tools used within the implementation of the quantum Jacobi solver can be applied to other classical iteration schemes. For example, we can split the matrix $R=B+T$ into a bottom off-diagonal matrix $B$ and top off-diagonal matrix $T$. In this case, the Gauss-Seidel scheme can be employed which has the following iterative scheme \cite{linear_systems_textbook},
\begin{equation} \label{eq:quantum_gauss_seidel}
    \vb*{x}_k=(D+B)^{-1}(\vb*{b}-T \vb*{x}_{k-1})  \ \forall \ k\in[1,K].
\end{equation}
While Eq.~\eqref{eq:quantum_gauss_seidel} does not readily map to quantum operators, we propose to use the Woodbury identity \cite{woodbury} and recast the iteration scheme using the recursive summation. The Gauss-Seidel iteration then becomes
\begin{equation}
    \vb*{x}_k= \left\{\sum_{l=0}^{\infty} (-D^{-1}B)^l \right\} D^{-1}(\vb*{b}-T\vb*{x}_{k-1}) \approx \Omega D^{-1}(\vb*{b}-T\vb*{x}_{k-1}),
\end{equation}
with a Woodbury summation term $\Omega \eqbydef \sum_{l=0}^L (-D^{-1}B)^l$ being truncated to some integer-valued order $L$. Generally, we note that a modest truncation order $L$ provides excellent results in practice. We provide more information regarding the convergence with $L$ and numerical tests in the Appendix.

\section{III. Implementation}

We proceed to describe an implementation of quantum iterative solvers on a quantum processor with the help of block encoding techniques. Specifically, we concentrate on the quantum Jacobi solver. To execute the iteration presented in Eq.~\eqref{eq:quantum_jacobi}, an efficient method for statevector subtraction is needed. Implementing this in practice involves devising a circuit capable of subtracting the amplitudes of two states, one of which is generally unknown as it encodes the prior iterate solution. The no-cloning theorem prohibits the creation of a deterministic circuit for this purpose \cite{no_quantum_adder}. Despite this, probabilistic statevector subtraction is achievable \cite{quantum_adder_with_lcu}. One way of implementing iterative statevector subtraction is to make use of the linear combination of unitaries approach \cite{hamiltonian_simulation_lcu}. In particular, we suggest to use a \ac{LCU} for preparing the superposition of multiple states representing all parts of the iterate state at a given step $k$.
\begin{figure*}[t]
    \centering
    \includegraphics[width=0.9\linewidth]{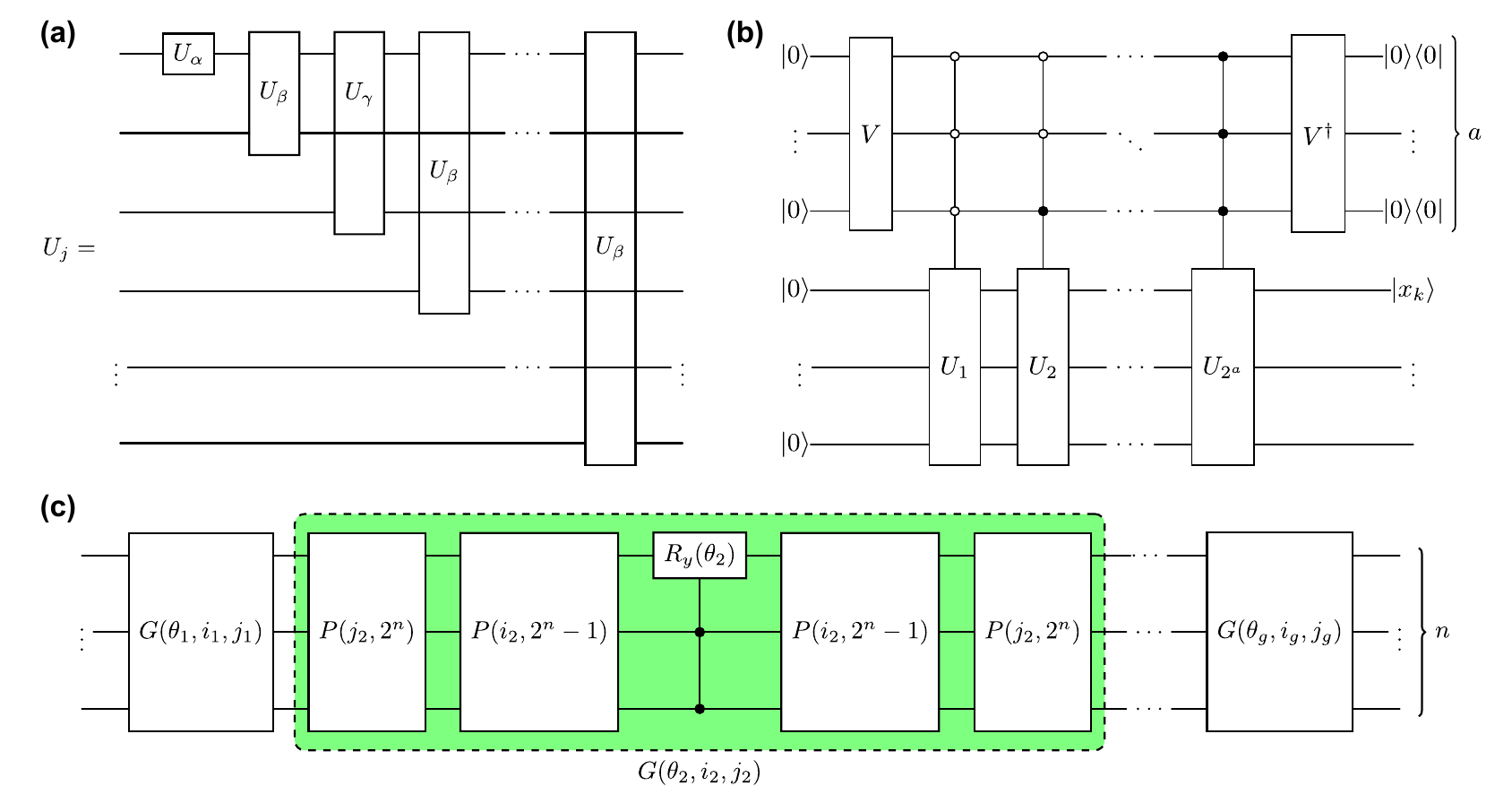}
    \caption{The quantum circuit used to implement the Jacobi iterative method. \textbf{(a)} The multiplication circuit used to obtain the block encoding $U_j$ of each expansion term. The block encoding $U_\alpha$ represents the vector encodings $U_b$ or $U_0$. The block encodings $U_\beta$ and $U_\gamma$ represent the matrix encodings $U_{D^{-1}}$ and $U_R$. After multiplying the block encodings, the top $\log_2(N)$ qubits encodes the $j^\text{th}$ expansion term. \textbf{(b)} The \ac{LCU} circuit used to solve for the $k^\text{th}$ iterate $\ket{x_k}$ using $a=\lceil\log_2(k+1)\rceil$ ancillary qubits. The unitaries $U_j$ represent the block encodings of the $j^\text{th}\in[1,2^a]$ expansion term. The unitary $V$ encodes the normalization constants associated with each block encoded expansion term. The final result is obtained after taking a projective measurement on the ancillary qubits. \textbf{(c)} The block encoding circuit used to encode matrices $D^{-1}$ and $R$ and vectors $\vb*{x}_0$, $\vb*{b}$ and $\vb*{v}$. The circuit is constructed from a product of $g$ Givens rotations over $n$ qubits. Each Givens rotation $G(\theta,i,j)$ consists of permutation gates $P(j,2^n)$, $P(i,2^n-1)$ and a $y$-axis rotation by angle $\theta$ that is controlled by $n-1$ qubits denoted $R_y(\theta)$.}
    \label{fig:circuits}
\end{figure*}

Let us consider the quantum iteration scheme in Eq.~\eqref{eq:quantum_jacobi} and expand the recursion such that the $k^\text{th}$ iterate is composed of the linear combination of operators acting on the computational zero state. 
The equation can be written as an expansion of the form
\begin{equation} \label{eq:jacobi_expansion}
    \ket{x_k} = \Biggl[\sum_{j=1}^k (-U_{D^{-1}}U_R)^{j-1} U_{D^{-1}}U_b +(-U_{D^{-1}}U_R)^k U_0 \Biggr]\ket{\emptyset},
\end{equation}
consisting of $k+1$ terms. We can write this in the LCU form as $\ket{x_k} \eqbydef \mathcal{U}\ket{\emptyset}$. Each term in the expansion can be represented as a unitary by multiplying the individual block encodings of the known quantities. The circuit used to perform this multiplication is given in Fig.~\ref{fig:circuits}(a). The circuit is constructed such that the individual block encoding of each matrix is multiplied with the use of ancillary qubits \cite{algorithms_survey}. This is a necessity because the matrices are individually embedded within larger block matrices to form unitary operators. The nature of this embedding means that the direct multiplication of the block encodings introduces additional unwanted cross terms. These terms arise from the multiplication of the off-diagonal blocks within the individual embeddings. These additional terms interfere with the required data structure, which is expected to only encode the data corresponding to the multiplication of the underlying matrices themselves. As shown in Fig.~\ref{fig:circuits}a, this is resolved by extending the register such that block encodings are only partially overlapped. 
After the multiplication, the Jacobi expansion takes the form of 
\begin{equation} \label{eq:lcu_jacobi}
    \ket{x_k} = \sum_{j=1}^{k+1}\pm c_jU_j\ket{\emptyset}. 
\end{equation}
The coefficients $c_j$ refer to the multiplication of the normalization constants associated with the data encodings. The top $\log_2(N)$ qubits encode the multiplied data corresponding to the $j^\text{th}$ expansion term. 

We also note that instead of multiplying the block encodings directly, it can be beneficial to encode the powers of the block encodings. For example, we see from Eq.~\eqref{eq:jacobi_expansion} that multiple powers of the term $U_{D^{-1}}U_R$ are required. Encoding matrix powers can be achieved naturally with a \ac{QSP} framework in order to reduce the quantum resource requirement in performing block encoding multiplications. More details regarding overall resource requirements are explained in Sec. V. 

\par To account for the coefficients $c_j$, it is necessary to construct a normalization unitary \cite{hamiltonian_simulation_lcu}. Firstly, the normalization vector $\vb*{v}$ is calculated according to 
\begin{equation} \label{eq:normalisation_unitary}
    v_j = \frac{\sqrt{c_j}}{\sum_{j=1}^{k+1} c_j} \text{ , } c_j =     
    \begin{cases}
      \tilde{b} \tilde{r}^{j-1} \tilde{d}^j & \text{if $j<k+1$}\\
      \tilde{b}^j \tilde{d}^j \tilde{x}_0 & \text{if $j=k+1$}.
    \end{cases}
\end{equation}
The normalization factors are $\tilde{b}=\norm{\vb*{b}}$, $\tilde{x}_0=\norm{\vb*{x}_0}$, $\tilde{r}=\norm{R}$ and $\tilde{d}=||D^{-1}||$. Secondly, block encoding is applied to create the resulting unitary $V$ that encodes this information, which is defined by $\ket{v} \eqbydef V\ket{\emptyset}$.

\par Implementing the addition and subtraction of the expansion terms in Eq.~\eqref{eq:lcu_jacobi} can be achieved with the \ac{LCU} circuit shown in Fig.~\ref{fig:circuits}(b). The circuit accounts for all of the terms in the full expansion of the $k^\text{th}$ iterate solution using controlled versions of the multiplication unitaries that encode the problem data. Note that the minus signs in Eq.~\eqref{eq:lcu_jacobi} can be incorporated into unitaries $U_j$ by including additional phase gates. In order to prepare the approximate solution $\ket{x_k}$, a total of $a=\lceil\log_2(k+1)\rceil$ control qubits are required to embed $2^a$ controlled expansion unitaries. In the case where $2^a>k+1$, an additional $j = 2^a-k-1$ zero terms are appended to $\vb*{v}$. This corresponds to the calculation of the iterate solution where there are fewer terms in the Jacobi expansion compared to the number of controlled unitaries required by the \ac{LCU} circuit. The normalization unitary $V$ is therefore crucial in accounting for both the normalization of the encoded data and the calculation of iterate solutions where $k+1$ is not a power of two. Upon implementing the full \ac{LCU} circuit, the superposed solution $\ket{x_k}$ is obtained by projecting the auxiliary register onto the zero state.  

\par The final step in the quantum circuit implementation of the Jacobi method is to extract the relevant information from the solution that is prepared as a quantum state. Generally, this is not easy due to the readout problem, which represents one of the challenges for quantum linear algebra solvers \cite{Biamonte2017,Montanaro2016}. One option corresponds to accessing features of the solution by measuring an expectation $\bra{x_k} M \ket{x_k}$ of an observable that is represented by operator $M$. This observable can represent statistical moments of a field \cite{hhl}, a principal component \cite{Lloyd2014}, or correspond to evaluating the solution at a particular grid point \cite{paine2023physicsinformed}. Alternatively, shadow tomography can be used to measure the state $\ket{x_k}$ in the computational basis \cite{shadow_tomography}. After post-processing, several amplitudes of the classical solution vector $\vb*{x}_k$ can be measured. The final solution will then be an approximation to the true solution $\vb*{x}$ after both solutions have been normalized.

\par When solving for the $K^\text{th}$ iterate as an approximate solution of a differential equation, the quantum state $\ket{x_{K}}$ effectively embeds all of the prior iterate solutions that spans from $\ket{x_0}$ to $\ket{x_{K-1}}$. While these prior iterates are naturally stored in the history of the state, it is noted that there is no easy way to access $\ket{x_{k<K}}$ within the \ac{LCU} circuit itself without having to perform a series of mid-circuit measurements. Nonetheless, the circuit construction is valid for any $k$ such that any prior iterate solutions can be measured from the output of successive circuit implementations, if required.

\par The steps taken to implement the quantum analog of the Jacobi method can be applied to the Gauss-Seidel method. In this case, the quantum state solution at iteration $k$ is obtained from the full expansion when written in terms of quantum objects with block encodings $U_B, U_{D^{-1}}, U_T, U_b\text{ and }U_0$. Thus, the quantum Gauss-Seidel scheme from Eq.~\eqref{eq:quantum_gauss_seidel} can be written as an expansion of the form 
\begin{equation}
    \ket{x_k} = \Biggl[ \sum_{j=0}^{k-1} (-U_\Omega U_{D^{-1}}U_T)^j U_\Omega U_{D^{-1}}U_b + (-U_\Omega U_{D^{-1}}U_T)^k U_0 \Biggr] \ket{\emptyset}.
\end{equation}
The Woodbury summation $U_\Omega \eqbydef \sum_{l=0}^L (-U_{D^{-1}}U_B)^l$ can be implemented with an \ac{LCU} circuit, using techniques such as \ac{QSP} to embed the $l^\text{th}$ power of $U_{D^{-1}}U_B$. This circuit is then embedded within the $k+1$ terms of the expansion. These terms form the controlled unitaries of the \ac{LCU} circuit from Fig.~\ref{fig:circuits}(b), which is then measured to extract the solution state $\ket{x_k}$. 

Let us now discuss possible implementations for block encodings. First, we note that this can be done via methods of \ac{LCU}-based encoding or an oracle-based encoding \cite{Low2019hamiltonian}. Second, the approach can benefit from block encodings with sparse matrices and degenerate matrix elements, where efficient strategies based on oracles have been put forward recently \cite{block_encoding_1, block_encoding_2, block_encoding_3, block_encoding_4}. In the following we put forward another option where the use of oracles is avoided, and unitary representations of operators are developed with the help of Givens rotations \cite{Reck1994,Wecker2015,qr_decomposition_algo}. The Givens-based block encoding protocol and the corresponding quantum circuit are shown in Fig.~\ref{fig:circuits}(c). The procedure begins by embedding the data within a unitary matrix and then applying a QR decomposition to express the matrix as a product of Givens rotations \cite{qr_decomposition_algo}. Each Givens rotation is then implemented using $n$ qubits using the following gate sequence,
\begin{equation}
    G(\theta,i,j) = P(j,2^n)P(i,2^n-1) \text{C}^{\otimes (n-1)}\text{-}R_y(\theta) P(i,2^n-1)P(j,2^n).
\end{equation}
Here, $\text{C}^{\otimes (n-1)}\text{-}R_y(\theta)$ is a single qubit rotation by angle $\theta$ about the $y$-axis that is controlled by $n-1$ qubits and $P(i,j)$ are permutation gates. The permutation gates $P$ are matrices with components $P_{ij}=P_{ji}=1$, $P_{ii}=P_{jj}=0$ and $P_{h\neq i,j;h\neq i,j}=1$. These permutation gates can be decomposed into a series of CNOT and SWAP operations \cite{universal_givens}.

We note that the procedure used to embed the data using the Givens block encoding protocol depends on the data structure. To begin with, all of the known classical quantities are normalized prior to encoding. Their normalization constants are accounted for in the construction of unitary $V$ via Eq.~\eqref{eq:normalisation_unitary}, which is block encoded using $\lceil\log_2(k+1)\rceil$ qubits. To see how the block encoding is implemented in practice, consider a non-unitary matrix $W$ that has dimensions of $N\times N$. This matrix is first substituted into the unitary matrix $\widetilde{W}$ according to
\begin{equation}
    \widetilde{W} = \begin{pmatrix} W&\sqrt{\mathds{1}-W^\dagger W} \\ \sqrt{\mathds{1}-W^\dagger W}&-W \end{pmatrix}.
\end{equation}
The QR decomposition is applied to $\widetilde{W}$ to decompose it into a product of Givens rotations \cite{qr_decomposition_algo}. This matrix is then encoded into the gate sequence $U_W$ using the Givens block encoding circuit with $\lceil \log_2(2N) \rceil$ qubits. The full construction of the block encoding circuit can be simplified for instances where $W$ has a particular sparsity pattern. In the Jacobi scheme, $W$ takes the form of a purely diagonal matrix or a banded structure with zero diagonal components. In these cases, there are fewer Givens rotations in the QR decomposition and the resulting construction of the permutation matrices are simplified. The overall circuit depth however depends on the chosen finite differencing scheme and details of the problem.

Finally, returning back to block encodings required specifically by Jacobi iterations, we substitute $W$ by the relevant operators. The Givens-based block encoding circuit $U_R$ is obtained by encoding $R$ in the top-left matrix block of $\widetilde{W}$. For the diagonal part $D^{-1}$, the matrix inversion is first performed classically by simple inversion of elements (and assuming they are non-zero). This largely simplifies the workflow and allows avoiding expensive quantum operator inversion \cite{hhl}. The inverted matrix is then encoded into the top-left matrix block of $\widetilde{W}$ and the unitary $U_{D^{-1}}$ is prepared with the block encoding circuit. For vectors $\vb*{x}_0$ and $\vb*{b}$ of length $N$, their unitary embeddings $U_0$ and $U_b$ are implemented using the Givens block encoding circuit with $\log_2(N)$ qubits. This approach eliminates the need for extra ancillary qubits compared to traditional state preparation subroutines \cite{state_preparation}.
\begin{figure}
    \centering
    \includegraphics[width=\linewidth]{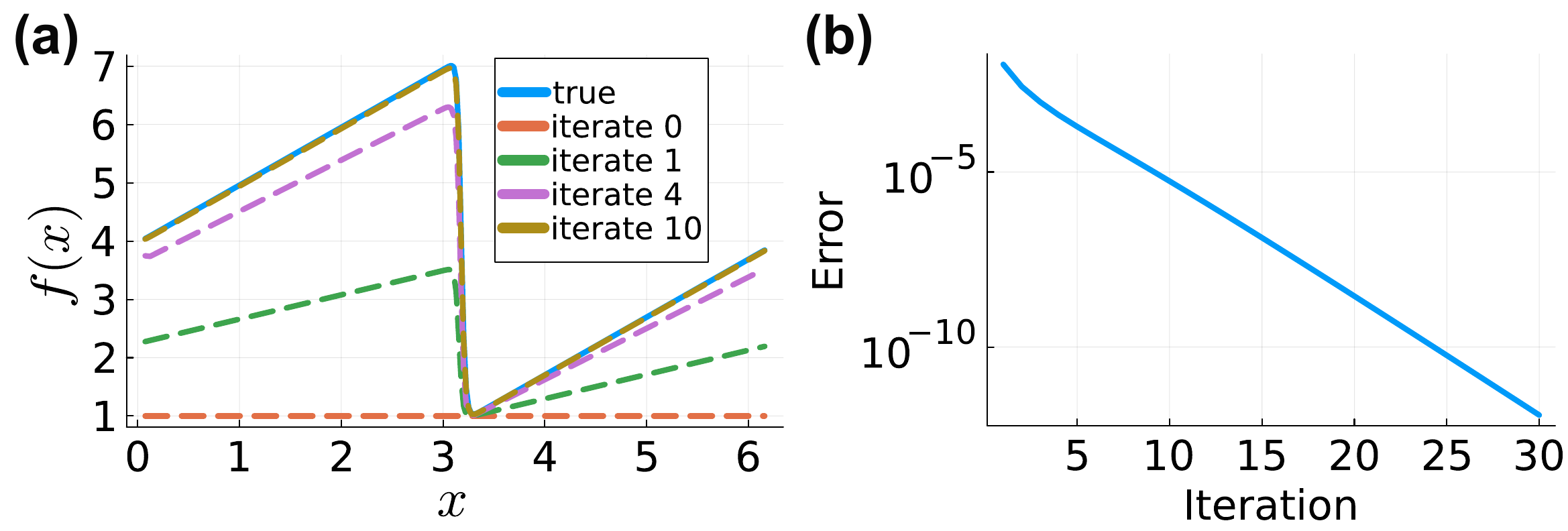}
    \caption{A demonstration of how the output from the quantum implementation of the Jacobi method iteratively converges towards the true solution. \textbf{(a)} The problem is defined by the viscous Burgers equation, whose true solution is a shock wave as displayed by the solid curve. The dashed curves represent the $k^\text{th}\in[0,1,4,10]$ iterate solution obtained from the quantum Jacobi solver. \textbf{(b)} The logarithmic decrease in the error of the quantum Jacobi solution over $30$ iterations.}
    \label{fig:demo}
\end{figure}

\section{IV. Results}

We proceed to address the application of quantum iterative solvers as applied to problems in computational fluid dynamics. One of the examples for \ac{CFD} challenges involves modeling convection-diffusion systems. The physics of these systems is dictated by the viscous Burgers equation. This equation is employed in various fields such as aerodynamics, biology and cosmology to study phenomena related to fluid mechanics, shock waves and turbulence \cite{burgers_litrev}. The viscous Burgers equation in one-dimension is expressed as
\begin{equation} \label{eq:burgers_equation}
    \frac{\partial f}{\partial t} + f\frac{\partial f}{\partial x} = \mu\frac{\partial^2f}{\partial x^2} \text{~~for~~} 0\leq x \leq L_x \text{~~and~~}0 \leq t \leq T,
\end{equation}
where $\mu$ represents the fluid viscosity. The boundary conditions and initial condition are given respectively by
\begin{equation}
f(0,t)=f_0,\text{~~}f(L_x,t)=f_{L_x}\text{~~and~~}f(x,0)=g(x).
\end{equation}
This \ac{PDE} models the scalar field $f(x,t)$ of a dissipative system across a space of length $L_x$ over time $T$ from a smooth initial condition function $g(x)$.

We now exemplify solving the Burgers equation with an iterative solver. The first step is to discretize the \ac{PDE} and obtain a system of equations. Using an appropriate finite differencing method, a mesh is defined which consists of $N+1$ spatial nodes uniformly spaced in the interval $[0,L_x]$ and $M+1$ temporal nodes uniformly spaced in the interval $[0,T]$. The quantum implementation of the Jacobi method is then utilized to perform the iterative matrix inversion of the resulting system of equations, and obtain a quantum state with amplitudes proportional to the scalar field solution of Eq.~\eqref{eq:burgers_equation}.

In Fig.~\ref{fig:demo} we illustrate how the iterate solutions converge during the quantum implementation of the Jacobi method. The result considers a snapshot solution to the viscous Burgers equation with $\mu=0.05$, at a fixed point in time. The initial and boundary conditions are defined by a shock wave profile. The iterates in Fig.~\ref{fig:demo}(a) demonstrate how the intermediate solutions obtained from the algorithm converges towards the true solution of the differential equation (here we show the full solution for illustrative purposes). The error scaling of the subsequent quantum solution is plotted in Fig.~\ref{fig:demo}(b) with the error taken to be one minus the fidelity. The fidelity is given by $\left\vert\bra{f(x)}\ket{f_k(x)}\right\vert^2$, where both the true solution $f(x)$ and the Jacobi solver solution $f_k(x)$ have been normalized beforehand. We observe that the accuracy of the result improves significantly in only a few iterations, even in cases like this where the solution contains a discontinuity in the form of a shock wave. 
\begin{figure}[t]
    \centering
    \includegraphics[width=0.85\linewidth]{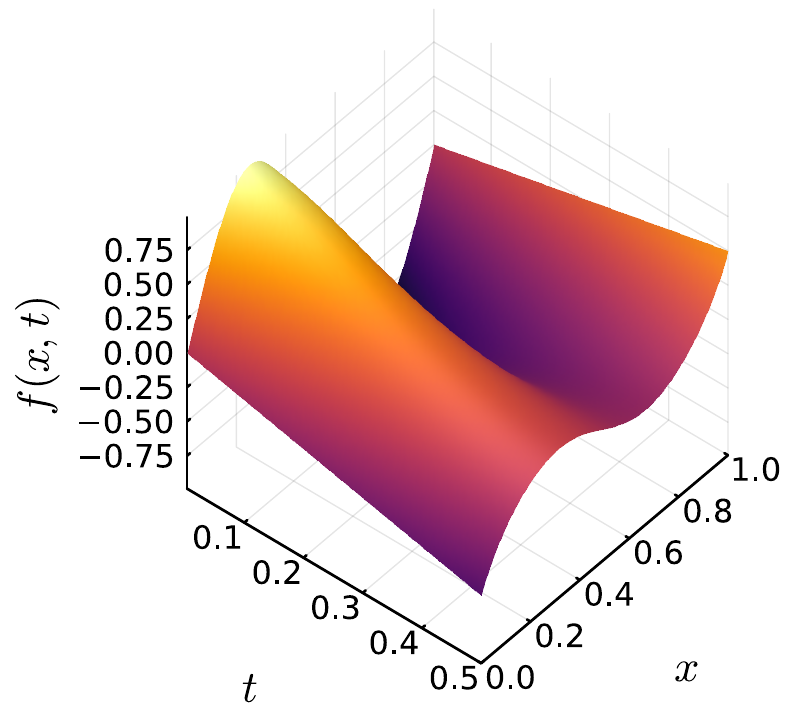}
    \caption{The solution to the Burgers equation using the quantum circuit implementation of the Jacobi method after $80$ iterations. The scalar field surface $f(x,t)$ represents a travelling sinusoidal wave in a dissipative system.}
    \label{fig:burgers_surface}
\end{figure}

\par An example solution to the Burgers equation using the quantum circuit implementation of the Jacobi method is shown in Fig.~\ref{fig:burgers_surface}. The problem is given by the Burgers equation defined in Eq.~\eqref{eq:burgers_equation} with $\mu=0.08$, $L_x=1$ and $T=0.5$ (using dimensionless units for brevity). The finite differencing mesh is defined with $N=128$ and $M=150$. Using a backward time centered space differencing scheme, the \ac{PDE} is converted into a system of equations of the form 
\begin{equation} \label{eq:fdm_system}
    f(x,t_m)=A^{-1}f(x,t_{m-1}) \ \forall \ m\in[0,M]. 
\end{equation}
Here, the matrix $A$ encapsulates the nonlinear differential components within a tridiagonal structure. The initial condition is set to $g(x)=\sin(2\pi x)$ with dynamic boundary conditions such that $f(0,t_m)=-t_m$ and $f(L_x,t_m)=t_m$. The surface plot visualizes the scalar field solution, which represents a wave travelling through a medium with non-zero viscosity.

\par The solution to the system of equations is obtained from the simulation of the quantum Jacobi algorithm outlined in Fig.~\ref{fig:pipeline} with $k=80$. The numerical simulation of quantum circuits is performed using Julia's Yao package \cite{yao}, including full circuit decompositions. We observe that high-quality solutions can be obtained from the full statevector using limited resources.

Next, we proceed to another example in the \ac{CFD} domain. The study of aeroacoustics considers the interaction between noise generation and fluid flow. Accurately modeling the propagation of sound waves from the perspective of aerodynamic flows is a challenge that concerns the aerospace and automotive sectors \cite{airbus_bmw}. The fluid dynamics is governed by the Navier-Stokes equations which reduces to the Euler equations in the case of an inviscid flow. For acoustic fields, the problem reduces further to the study of the linearized Euler equations \cite{aeroacoustic}. The two-dimensional linearized Euler equations for a quiescent fluid with negligible base flows are given by 
\begin{align} \label{eq:euler_equations}
    \frac{\partial p}{\partial t} + \bar{\rho} \bigg( \frac{\partial u}{\partial x} + \frac{\partial v}{\partial y} \bigg) = 0, \\
    \frac{\partial u}{\partial t} + \frac{1}{\bar{\rho}} \frac{\partial p}{\partial x} = 0, \\
    \frac{\partial v}{\partial t} + \frac{1}{\bar{\rho}} \frac{\partial p}{\partial y} = 0.
\end{align}
Here, $u(x,t)$ is the velocity component in the $x$-direction, $v(y,t)$ is the velocity component in the $y$-direction and $\bar{\rho}$ is the mean density of the fluid. These equations model the propagation of acoustic waves via the pressure field solution $p(x,y,t)$ in quiescent air by considering the conservation of mass and momentum.  
\begin{figure}[t]
    \centering
    \includegraphics[width=0.6\linewidth]{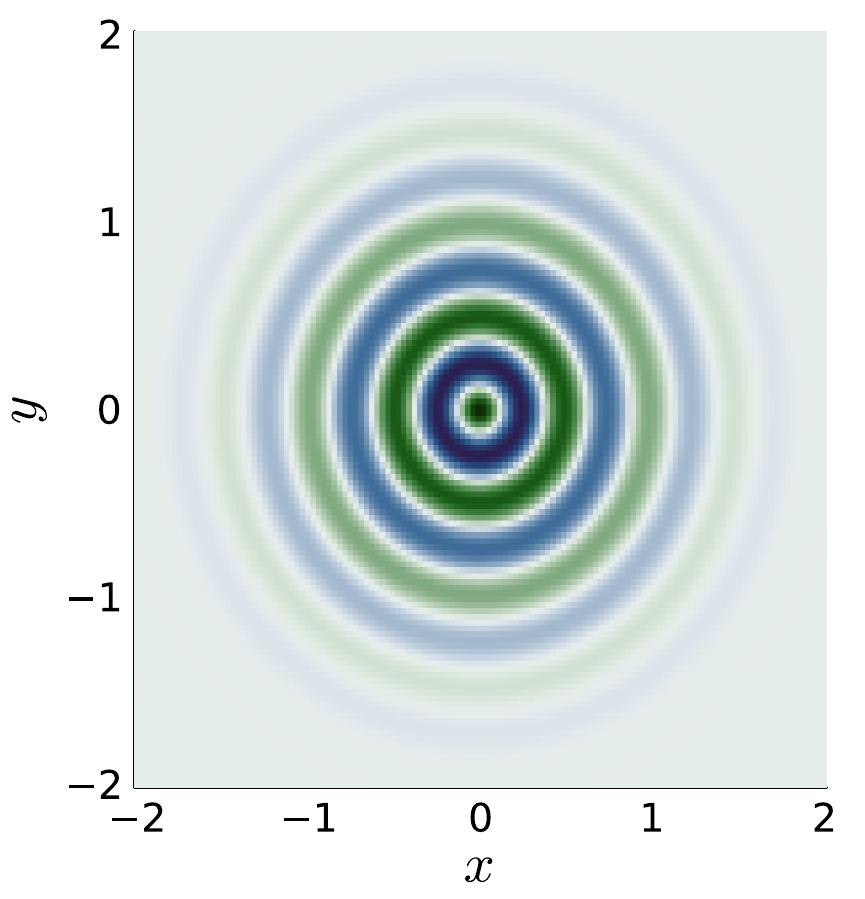}
    \caption{The solution to the Euler equations using the quantum circuit implementation of the Jacobi method after $12$ iterations. The pressure field solution $p(x,y)$ represents the transmission of a Gaussian-modulated noise, emitted from a static point source, through quiescent air.} 
    \label{fig:pressure_field}
\end{figure}

\par An example solution to the Euler equations using the quantum circuit implementation of the Jacobi method is shown in Fig.~\ref{fig:pressure_field}. This solution uses a Gaussian-modulated sinusoidal initial condition $\cos(2\pi\omega r)e^{-r^2}$ with frequency $\omega=2$ and radial component $r^2=x^2+y^2$. The solution also uses non-reflective boundary conditions. As before, the solution is obtained firstly by performing backward time centred space differencing on the two-dimensional \ac{PDE} system. The discretization is applied in the $x$-direction using $N_x=128$ spatial nodes with $x\in[-2,2]$ and in the $y$-direction using $N_y=128$ spatial nodes with $y\in[-2,2]$. The temporal domain is divided into $M=60$ temporal nodes. The solution is then obtained from the simulation of the quantum Jacobi algorithm with $k=12$. The resulting contour plot visualizes the pressure field solution of the noise transmitted from a static point source. From this, we see how the Gaussian-modulated sinusoidal waves are travelling through a quiescent fluid. 

\begin{figure}
    \centering
    \includegraphics[width=0.72\linewidth]{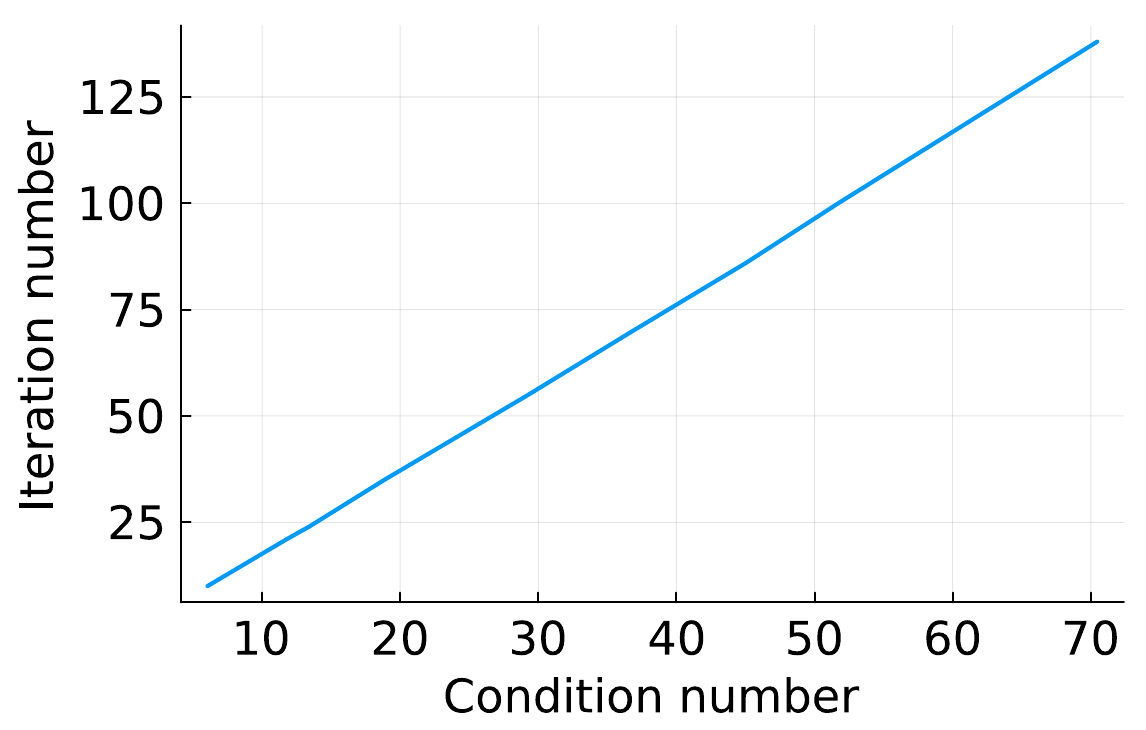}
    \caption{The scaling of the number of iterations with condition number for a system of fixed size. The iteration number represents the minimum number of iterations required such that the error between the true solution and the quantum Jacobi solution meets a given threshold.}
    \label{fig:condition_number}
\end{figure}

Finally, we assess the complexity of running the iterative algorithms based on linear systems of equations properties, namely the condition number. The scaling behaviour of the quantum circuit implementation of the Jacobi method with respect to condition number $\kappa$ is shown in Fig.~\ref{fig:condition_number}. A tridiagonal system of dimension $256\times 256$ is solved using the quantum Jacobi solver with different numerical values. The respective condition numbers of these systems ranges from $3$ to $70$. Their respective iteration numbers correspond to the minimum number of iterations such that the error between the true solution and the quantum Jacobi solution meets a threshold of $10^{-6}$. Here, the true solution is taken to be the direct inversion of the resulting matrix $A$. The cost associated with performing iterations scales linearly with the condition number, suggesting that systems with higher condition numbers require more iterations to converge towards a solution with a given level of accuracy. It should be noted that the cost of each iteration is independent of the condition number. This scaling of iterations $K\sim\mathcal{O}(\kappa)$ is representative of the depth of the \ac{LCU} circuit and confirms the observation that the optimal scaling of a linear systems quantum solver with respect to condition number is strictly linear \cite{adiabatic_random_walk}. 

\section{V. Discussion}

\par The total circuit width for the quantum Jacobi circuit accounts for the qubits required to multiply the block encoded data and the control qubits required to perform the statevector addition and subtraction. As a result, the circuit requires $\log_2(N)+2k+\lceil\log_2(k+1)\rceil$ qubits. The linear dependence on the iteration number $k$ stems from the multiplication circuit from Fig.~\ref{fig:circuits}(a). Where required it is possible to remove this dependence by using \ac{QSP} in place of the existing multiplication strategy. \ac{QSP} then uses at most two additional ancillary qubits to apply polynomial transformations to block encoded matrices \cite{qsp}. To see why \ac{QSP} may be useful, consider writing the expansion of the Jacobi scheme from Eq.~\eqref{eq:classical_jacobi} in terms of matrix $Q=D^{-1}R$ and vector $\vb*{p}=D^{-1}\vb*{b}$. After respectively block encoding the data into unitaries $U_Q$ and $U_p$, Eq.~\eqref{eq:jacobi_expansion} can be rewritten as
\begin{equation}
    \ket{x_k} = \Biggl[\sum_{j=1}^k(-U_Q)^{j-1}U_p + (-U_Q)^kU_0  \Biggr]\ket{\emptyset}.
\end{equation}
The encoded power of $U_Q$ can then be directly multiplied by $U_p$ or $U_0$ without requiring any additional qubits. With \ac{QSP} the total circuit width becomes $\log_2(N)+4+\lceil\log_2(k+1)\rceil$, which now has only logarithmic dependence on iteration number. 

\par The total circuit depth for the quantum Jacobi circuit accounts for the number of gates associated with the block encoding of the data and the operations used to manipulate the subsequent unitaries. The dominant contribution in terms of gate count comes from the $\mathcal{O}(k^2)$ multiplications of the block-encoded matrices. This gate count scaling is also applicable for \ac{QSP} since the depth of the circuit for terms $(-U_Q)^d$ are proportional to the degree $d$ of the polynomial \cite{fault_tolerant_qsp}. Generally speaking, the choice of block encoding procedure used to embed the data should be dependent on problem specifics such as the number of repetitions and the number of non-zero matrix elements of $A$. If the cost associated with the chosen block encoding protocol is $C$, the total circuit depth will be $\mathcal{O}(k^2C)$. To date the research on identifying optimal block encoding protocols \cite{block_encoding_1, block_encoding_2, block_encoding_3, block_encoding_4} points that for a given problem, its cost typically depends on the sparsity $s$ and multiplicity $z$ of the matrix values such that $C = C(s,z)$. This circuit implementation is thus most favourable for sparse systems, where the block encoding of the problem data is efficient and the resulting circuit depth is reduced.

\par The accuracy, reliability and efficiency of numerical techniques play a crucial role in obtaining dependable results in \ac{CFD}. In terms of computational savings, the quantum circuit implementation of the Jacobi method offers a significant enhancement in data storage efficiency when compared to its classical counterpart. An exponential speed-up in memory comes from the fact that optimized classical iterative approaches achieve a scaling of $\mathcal{O}(N)$, while the quantum Jacobi solver has an improved scaling of $\mathcal{O}(\log_2(N))$ (of course, keeping in mind that the solution is stored as a quantum state). This is a characteristic that is particularly relevant for extensive industrial \ac{CFD} simulations where the memory required to store and manipulate the data exceeds the amount available in classical computing. In addition, the convergence rate is the same as the classical implementation whereby convergence is obtained in the same number of iterations. The quantum Jacobi solver therefore offers promising prospects for reducing the computational resources required for industrial simulations, especially when embedded within higher-dimensional multigrid algorithms. 

In discussing the role of quantum iterative solvers for PDEs and \ac{CFD} problems, we note that they provide an alternative approach that is not covered by other options. Approaches based on variational state preparation are frugal in terms of qubit numbers and depth, but are largely dependent on the ability to train state preparation circuits \cite{vqas}. On the other hand, approaches like the \ac{QLSA} are geared towards large-scale fault tolerant quantum processors, where the direct matrix inversion can be performed assuming access to a large number of ancillas \cite{hhl}. In contrast, quantum iterative solvers fall in between these two regimes and are geared towards early fault tolerant devices. Due to the iterative nature of these algorithms, fewer error corrected qubits are required and the subroutines used are less intensive. Furthermore, approaches like the quantum physics-informed neural networks \cite{kyriienko2021solving,paine2023physicsinformed,paine2023quantum} can be integrated with iterative solvers to reduce the number of iterations required.  

\par From an industrial \ac{CFD} perspective, approaches like the quantum Jacobi algorithm are best used as a preconditioning subroutine within larger classical multigrid algorithms. In this sense, only the most computationally intensive aspects of the simulations are off-loaded to a quantum device. These computational pipelines are a necessity for future surrogate modeling and digital twin simulations in leveraging the computational advantage of quantum physics. Machine learning approaches can also be used here to further increase the accuracy and performance of the quantum iterative solver \cite{iterative_solvers_with_ml}.

\section{VI. Conclusion}

We proposed algorithms for implementing quantum iterative solvers of differential equations based on a linear combination of unitaries represented by block encodings. Specifically, we demonstrated the building blocks for the quantum Jacobi iterative solver and developed a pipeline for preparing the approximate solution as a quantum state. This included the circuits for block encoding problems (differential equations), the construction of the circuit using the \ac{LCU} approach and the validation of the results in solving \acp{PDE}. The tools and techniques developed are readily applied to other iterative schemes, as demonstrated here with the extension to the Gauss-Seidel method where we have developed a particular expansion based on the Woodbury identity.

We have tested the approach by preparing solutions of the convection-diffusion systems described by the viscous Burgers equation, as well as sound wave propagation from the aerospace and automotive sectors which are industrially relevant \cite{airbus_bmw}. The results showed that high-quality solutions can be obtained already with limited resources. We note that the presented quantum Jacobi solver is best applied to problems where the solution contains an instability or discontinuity, advancing on techniques that require variational state preparation. The main advantage of using a quantum iterative solver over a classical iterative solver stems from the memory savings associated with manipulating the problem data and calculating the iterate solutions on quantum processors. This can be further improved with developing readout techniques and multigrid approaches.


\section{Acknowledgement}

We thank Konstantinos Agathos for illuminating discussions on the subject. This work was supported by Siemens Industry Software NV.

\appendix

\section{APPENDIX}

\par
\begin{figure}
    \centering
    \includegraphics[width=0.72\linewidth]{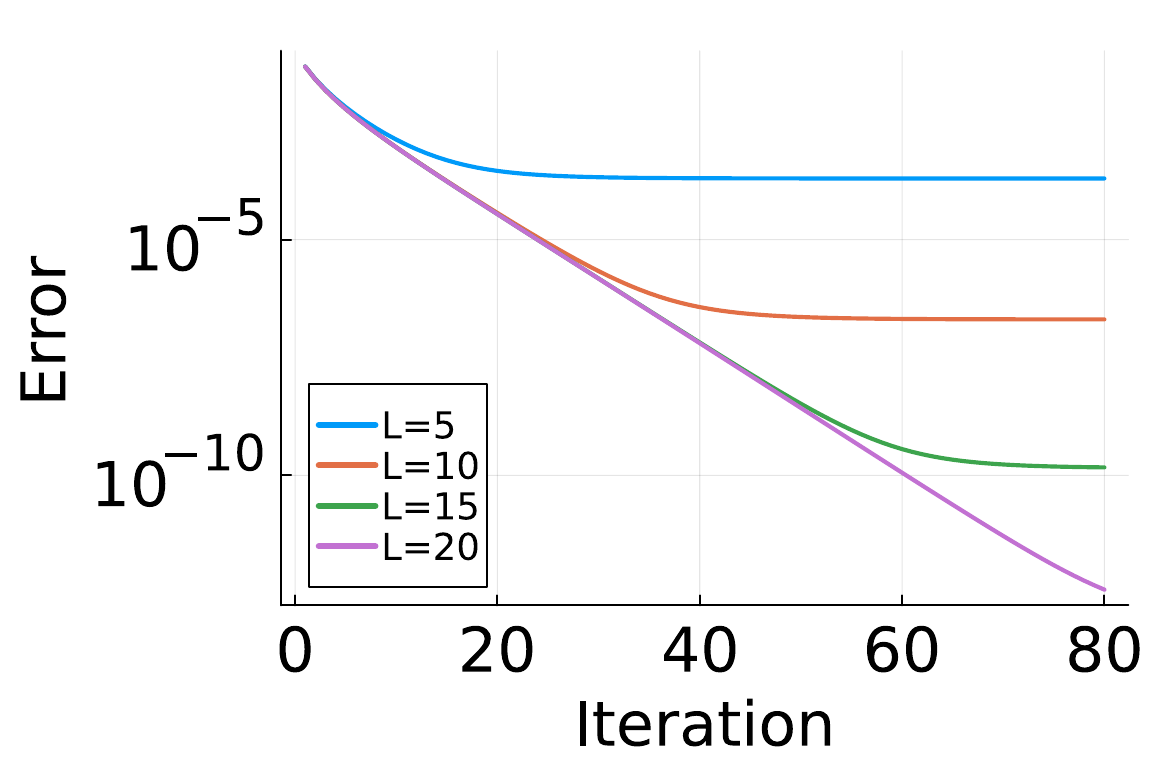}
    \caption{The logarithmic decrease in the error of the Gauss-Seidel solver over $80$ iterations for different values of the Woodbury summation truncation parameter $L\in[5,10,15,20]$.}
    \label{fig:woodbury}
\end{figure}

 The Gauss-Seidel iteration scheme can be written in terms of a truncated Woodbury summation using the Woodbury identity \cite{woodbury}. The influence that the truncation parameter $L$ can have on the overall error scaling of the Gauss-Seidel solver is demonstrated in Fig.~\ref{fig:woodbury}. This result considers solving a linear system of size $N=128$ with condition number $\kappa=24$.

\par The error reduces when $L$ is larger but this is at the expense of increased terms in the Woodbury summation. This translates to requiring more ancillary qubits in the \ac{LCU} circuit of the quantum implementation of the Woodbury term $U_\Omega$. As a consequence, the choice of truncation $L$ is dependent on a trade-off between improved convergence and increased quantum resources.


%


\end{document}